\documentclass{JHEP3}

\usepackage{amsmath,amsfonts,amsthm,times}

{\count255=\time\divide\count255 by 60
\xdef\hourmin{\number\count255}
\multiply\count255 by-60\advance\count255 by\time
\xdef\hourmin{\hourmin:\ifnum\count255<10 0\fi\the\count255}}
\def\ps@draft{\let\@mkboth\@gobbletwo
    \def\@oddhead{}
    \def\@oddfoot{\hbox to 7 cm{\tiny \versionno
       \hfil}\hskip -7cm\hfil\rm\thepage \hfil {\tiny\draftdate}}
    \def\@evenhead{}\let\@evenfoot\@oddfoot}
\def\draftdate{\number\month/\number\day/\number\year\ \ \ \hourmin }

\long\def\query#1{\hskip 0pt{\vadjust{\everypar={}\small\vtop to 0pt{\hbox{}%
     \vskip -13pt\rlap{\hbox to 49.0pc{\hfil{\vtop{\hsize=8pc\tolerance=6000%
     \hfuzz=.5pc\rightskip=0pt plus 3em\noindent#1}}}}\vss}}}}%
\newcommand\version[1] {\typeout{}\typeout{#1}\typeout{}
                   \vskip3mm \centerline{\fbox{{\tt DRAFT -- #1 -- }
                   {\small\draftdate}}} \vskip3mm}

\def\start{ 
\usepackage{amssymb,amsfonts,amsmath,amscd,latexsym,epsf}    
                 \setlength{\textwidth}{17cm} \setlength{\textheight}{25.2cm}
 \hoffset -20mm \topmargin= -13mm \begin{document}\version\versionno
}

\def \ignore#1 { {} } 

\def \Fig#1#2#3 {
\begin{figure}
\begin{center}
\scalebox{#2}{\includegraphics{#1.eps}}
\label{#1}
\end{center}
\caption{#3}
\end{figure}
}

\def \FF { {} _2{\cal F}_1 } 

\def \sp   { \ \ \ , \ \ \ }

\def \foint {\frac{1}{2i\pi}\oint }
\def \infint  {\int_{-\infty}^\infty }

\def \tsum  { {\textstyle \sum} }

\def \half {\frac{1}{2}}
\def \halfpi {\frac{\pi}{2}}
\def \halfi {\frac{i}{2}}
\def \halfib {\frac{i}{2b^2}}
\def \halfb  {\frac{b}{2}}
\def \ip {i\frac{\pi}{2}}
\def \hib {\frac{1}{2b}}
\def \hiib {\frac{i}{2b}}

\def \ap {\alpha'}
\def \ab {\bar{\alpha}}

\def \p {\partial}
\def \pp#1 {{\frac{\p}{\p #1}}}
\def \ppd#1 {{\frac{\p^2}{\p #1 ^2}} }

\def \om {\omega}
\def \g {\gamma}
\def \bg {\bar{\gamma}}
\def \vf {\varphi}
\def \s {\sigma}
\def \G {\Gamma}  
\def \e {\epsilon}
\def \a {\alpha}
\def \Ga {\Gamma_b}
\def \up {\Upsilon_b}
\def \t {\theta}
\def \vt {\vartheta_1}
\def \G {\Gamma}
\def \FF { {} _2{\cal F}_1 }

\def \rar {\rightarrow}
\def \lrar {\leftrightarrow}

\def \j  {-\half+iP}

\def \bz {\bar{z}}
\def \bw {\bar{w}}
\def \bx {\bar{x}}
\def \bp {\bar{\partial}}
\def \bq {\bar{q}}
\def \bL {\bar{L}}
\def \bJ {\bar{J}}
\def \tJ {\tilde{J}}

\def \Tr {{\rm Tr }\ }
\def \id {{\rm id } }
\def \sgn{ {\rm sgn} }
\def \cst {{\rm constant}}

\def \Z {\mathbb{Z}}
\def \N {\mathbb{N}}
\def \R {\mathbb{R}}
\def \C {\mathbb{C}}
\def \S {\mathbb{S}}

\def \la {\left\langle}
\def \ra {\right\rangle}
\def \lla {\langle\langle}
\def \rra {\rangle\rangle}

\def \tq {\tilde{q}}
\def \tth {\tilde{\theta}}
\def \ttau {\tilde{\tau}}

\def \uq  {U_q(sl_2)}
\def \asl {\widehat{s\ell_2}}
\def \SLC    {SL(2,\C)}
\def \SLR    {SL(2,\R)}
\def \SLU    {SL(2,\R)/U(1)}
\def \H      {H_3^+}

\def \six#1#2#3#4#5#6{
 _{#1\, #2}\left[\begin{array}{cc} 
#5 & #4 \\ 
#6 & #3
\end{array}\right]}

\def \CG#1#2#3#4#5#6{ \left(\begin{array}{ccc} #1 & #2 & #3 \\ #4 & #5
    & #6 \end{array} \right) }

\def \vir { \ , \ }
\def \F32#1#2#3#4#5#6{{} _3F_2\left(\left.\begin{array}{c}#1 \vir #2 \vir
    #3 \\ #4\vir #5
    \end{array} \right| #6 \right) }

\def\vect#1#2{ \left( \begin{array}{c} #1 \\ #2  \end{array} \right)}

\def \bea {\begin{eqnarray}}
\def \eea {\end{eqnarray}}
\def \bee {\begin{eqnarray*}}
\def \eee {\end{eqnarray*}}
\def \nn  {\nonumber}
\def \sst {\scriptscriptstyle}
\def \begm {\begin{multline}}
\def \enm  {\end{multline}}
\def \bega {\begin{align}}
\def \ena  {\end{align}}

\newcommand {\citer}[2]{\begin{flushright}{\small{[\ #1,\ \tcm{\it
          #2}\ ]}}
\end{flushright}}

\newcommand {\slidetitle}[1]{ \large \begin{center} { #1} \\
  \end{center} \normalsize \vspace{0.8cm}  }

\newcommand {\tcb}[1]{#1}
\newcommand {\tcr}[1]{#1}
\newcommand {\tcg}[1]{#1}
\newcommand {\tcm}[1]{#1}
\newcommand {\tcc}[1]{#1}
\newcommand {\tcy}[1]{#1}
\newcommand {\tcnoir}[1]{#1}

\usepackage{graphics}\usepackage{epic}
\usepackage{pstricks,pst-coil,pst-node}

\def \mycoil#1#2{\pscoil[coilarm=0,linewidth=.05,coilaspect=0,coilwidth=.4,
coilheight=1.2](2.2,0) \rput[#1]{*0}(2.5,0){$#2$}}

\newlength{\floatlength}

\def \mybdy#1#2#3{\pcline[linewidth=.22](0,0)(#3,0)
\pssetlength{\floatlength}{#3} \pssetlength{\floatlength}{.5\floatlength}
\rput[#1]{*0}(\floatlength,0.3){$\scriptstyle #2$}}

\def \extline#1#2{\psline[linewidth=.1](2.2,0)\rput[#1]{*0}(2.5,0){$#2$}}

\def \intline#1#2#3{\pcline[linewidth=.1](0,0)(#3,0)
\pssetlength{\floatlength}{#3} \pssetlength{\floatlength}{.5\floatlength}
\rput[#1]{*0}(\floatlength,0.3){$#2$}}

\def \degfield#1#2{\psdots[dotscale=1.5,dotstyle=*](0,0)
\rput[#1]{*0}(0,0.4){$#2$}}
\def \genfield#1#2{\psdots[dotscale=1.8,dotstyle=+,dotangle=45](0,0)
\rput[#1]{*0}(0,0.4){$#2$}}

\def \blockvertical#1#2#3#4#5#6#7{
\pspicture[](-3,-3)(3,3)
\rput[l]{20}(0,2.5){\extline{l}{#1}}
\rput[l]{160}(0,2.5){\extline{r}{#2}}
\rput[l]{-20}(0,-2.5){\extline{l}{#3}}
\rput[l]{-160}(0,-2.5){\extline{r}{#4}}
\rput[l]{180}(2.2,1.5){\mycoil{r}{\scriptstyle #5}}
\rput[l]{180}(2.2,-1.5){\mycoil{r}{\scriptstyle #6}}
\rput[l]{90}(0,-2.5){\intline{r}{#7}{5}}
\endpspicture}

\def \blockhorizontal#1#2#3#4#5#6#7{
\pspicture[](-5,-3)(5,2.8)
\rput[l]{70}(2.5,0){\extline{l}{#3}}
\rput[l]{-70}(2.5,0){\extline{l}{#4}}
\rput[l]{110}(-2.5,0){\extline{r}{#1}}
\rput[l]{-110}(-2.5,0){\extline{r}{#2}}
\rput[l]{-90}(1.5,2.2){\mycoil{t}{\scriptstyle #6}}
\rput[l]{-90}(-1.5,2.2){\mycoil{t}{\scriptstyle #5}}
\rput[l]{180}(2.5,0){\intline{t}{#7}{5}}
\endpspicture}

\def \blockhorh#1#2#3#4#5#6#7{
\pspicture[](-5,-3)(5,2.8)
\rput[l]{70}(2.5,0){\extline{l}{#3}}
\rput[l]{-70}(2.5,0){\extline{l}{#4}}
\rput[l]{110}(-2.5,0){\extline{r}{#1}}
\rput[l]{-110}(-2.5,0){\extline{r}{#2}}
\rput[br]{0}(2.5,.3){$\scriptstyle #6$}
\rput[bl]{0}(-2.5,.3){$\scriptstyle #5$}
\rput[l]{180}(2.5,0){\intline{t}{#7}{5}}
\endpspicture}

\def \blockverh#1#2#3#4#5{
\pspicture[](-3,-3)(3,3)
\rput[l]{20}(0,2.5){\extline{l}{#1}}
\rput[l]{160}(0,2.5){\extline{r}{#2}}
\rput[l]{-20}(0,-2.5){\extline{l}{#3}}
\rput[l]{-160}(0,-2.5){\extline{r}{#4}}
\rput[l]{90}(0,-2.5){\intline{r}{#5}{5}}
\endpspicture}

\newtheorem{claim}{Claim}
\newtheorem{hypothesis}{Hypothesis}

\title{Knizhnik--Zamolodchikov equations and spectral flow in 
  $AdS_3$ string theory}

\author{ Sylvain Ribault
\vspace{5mm}
\\
{\it  King's College London
\\ Department of Mathematics
\\ Strand, London WC2R 2LS
\\ United Kingdom }
\\ {\tt ribault@mth.kcl.ac.uk }
}

\abstract{
I generalize the Knizhnik--Zamolodchikov equations to correlators of spectral
flowed fields in $AdS_3$ string theory. If spectral flow is preserved
or violated by one unit, the resulting equations are equivalent to the
KZ equations. If spectral flow is violated by two units or more, only
some linear combinations of the KZ equations hold, but extra equations
appear. Then I explicitly show how these correlators and the associated conformal
blocks are related to Liouville theory correlators and conformal
blocks with degenerate field insertions, where each unit of spectral flow
violation removes one degenerate field. A similar relation to
Liouville theory holds for noncompact parafermions. 
}

\preprint{ 
\hepth{0507114}\\ KCL-MTH-05-07 \\
 }

\begin{document}

\section{Introduction and overview}

The Knizhnik--Zamolodchikov equations \cite{kz84} are an essential tool in the
study of conformal field theories with affine Lie algebra
symmetry\footnote{For 
a review of original work on such theories, see
  \cite{hkoc95} and references therein. In particular, early works on
  spectral flow are referenced in section 2.1 of that article.}. 
All correlation functions of affine primary fields obey this
system of linear differential equations, which determine their
dependence on worldsheet coordinates.

However, in string theory in $AdS_3$, whose 
associated conformal field theory has an affine $\SLR$ symmetry,
Maldacena and Ooguri have shown that the physical spectrum
cannot be built only from affine primaries \cite{mo00a}. Instead, one
should also include spectral-flowed fields. Correlation functions
involving such fields are not expected to obey the KZ equations.

Nevertheless, such
spectral-flowed fields are obtained from affine primaries via the
spectral flow automorphism of the affine Lie algebra. This will enable
me to derive generalized KZ equations for their correlation
functions. The other main purpose of this work is to explicitly relate
these correlation functions to Liouville theory correlations
functions. This amounts to solving the generalized KZ 
equations in terms of Virasoro conformal blocks.

Let me now sketch the results.
If a correlation function
respects spectral flow conservation, then it will satisfy a system of
equations (\ref{kzw}) which turns out to be equivalent to the KZ equations via a
simple twist\footnote{Such twisted KZ equations are already known in
  the context of WZW orbifolds \cite{bho01}.}
. (Actually, this conclusion also holds if spectral flow
conservation is violated by one unit, due to the global group
symmetry.) If a correlation function violates spectral flow
conservation, then it will satisfy only some specific linear
combinations (\ref{kzc}) of the KZ equations. However, the missing equations will
be replaced with simpler constraints (\ref{rpu}) which do not involve 
derivatives wrt worldsheet coordinates.

Therefore, the equations obeyed in the case when spectral flow is not
conserved are in some sense simpler than the original KZ
equations. This will become clear after I show how to perform
Sklyanin's separation of variables for such equations. Each unit of
spectral flow violation leads to the disappearance of one
variable-separated equation, until there are none left in the case of
maximal violation.

These variable-separated equations are actually identical to
Belavin--Polyakov--Zamolodchikov equations (\ref{bpz}). I will exploit this in
order to derive a relation between correlation functions of
$n$ spectral-flowed fields in the $\H$ model and correlation functions in Liouville
theory. (The $\H$ model, or string theory in the Euclidean $AdS_3$, is
introduced here for technical reasons.) If spectral flow conservation
is violated by $r$ units, the relevant Liouville correlation functions
will have $n-2-r$ degenerate field insertions (\ref{main}). This
shows why the maximal spectral flow violation $n-2$ is equal to the
number of Liouville degenerate fields needed to reproduce an unflowed
$\H$ correlator.
A similar relation with
Liouville theory also
holds for the $\SLU$ coset model (\ref{mmmm}).   

Deriving such relations between correlation functions involves not
only the KZ equations, but also the structure constants of the $\H$
model. The ordinary, spectral flow-preserving structure constant is
known to be equal to the Liouville structure constant associated with 
a Liouville vertex dressed with one degenerate field 
\cite{rt05}. Here I will show that the $\H$ spectral flow-violating
vertex corresponds to an ordinary Liouville vertex, at the levels
of structure constants (\ref{fpv}) (\ref{fpp}), operator product
expansions (\ref{opep}) (\ref{opev}), and conformal
blocks (\ref{cbl}). In particular, I will argue for the existence of a
spectral flow-violating operator product expansion in the $\H$ model,
as an alternative to the ordinary operator product expansion. This is
shown schematically in the diagrams below.
\bea
\begin{array}{r|l}
{\rm Euclidean}\ AdS_3\ \ \ \ \  & \ \ \ \ \ {\rm Liouville\ theory}
\\
\hline
\begin{array}{c} 
{\rm Flow-preserving} \\ {\rm vertex}
\end{array} 
\psset{unit=.3cm}
\pspicture[](-3,-2.5)(3,2.5)
\rput[l]{120}(0,0){\extline{l}{}}
\rput[l]{-120}(0,0){\extline{l}{}}
\rput[l]{0}(0,0){\extline{l}{}}
\endpspicture &
\psset{unit=.3cm}
\pspicture[](-2,-2.5)(4.5,2.5)
\rput[l]{120}(0,0){\extline{l}{}}
\rput[l]{-120}(0,0){\extline{l}{}}
\rput[l]{0}(0,0){\intline{l}{}{3}}
\rput[l]{90}(1,0){\mycoil{t}{}}
\endpspicture
\begin{array}{c}
{\rm Vertex\ dressed}\\{\rm with\ one\ degenerate\ field}
\end{array}
\\
\hline
\begin{array}{c} 
{\rm Flow-violating} \\ {\rm vertex}
\end{array} 
\psset{unit=.3cm}
\pspicture[](-3,-2.5)(3,2.5)
\rput[l]{120}(0,0){\extline{l}{}}
\rput[l]{-120}(0,0){\extline{l}{}}
\rput[l]{0}(0,0){\extline{l}{}}
\psdots[dotstyle=o,dotscale=2](0,0)
\endpspicture &
\psset{unit=.3cm}
\pspicture[](-2,-2.5)(8.5,2.5)
\rput[l]{120}(0,0){\extline{l}{}}
\rput[l]{-120}(0,0){\extline{l}{}}
\rput[l]{0}(0,0){\extline{l}{}}
\endpspicture
\begin{array}{c}
{\rm Ordinary}\\{\rm vertex}
\end{array}
\end{array}
\eea
In an Outlook, I will mention possible applications of these results
to string theory in $AdS_3$ and to the definition of a fusing matrix
for the $\H$ model.

\section{KZ equations and spectral flow \label{seckz}} 

In this section I derive which modifications of the KZ equations apply
to correlation functions involving spectral flowed fields.

\subsection{Preliminaries: definition of the spectral flowed fields}

Let me consider a conformal field theory with an chiral affine Lie
algebra symmetry $\asl$ at level $k>2$, living on the Riemann sphere parametrized by
complex coordinates $z,\bar{z}$. In this section, I will be
concerned only with the holomorphic sector. The
symmetry of this sector is a ``left-moving'' copy of the algebra $\asl$,
\bea
\left\{ \begin{array}{l}
[J^3_n,J^3_m]=-\frac{k}{2}n\delta_{n+m,0},\\ {} [J^3_n,J^\pm_m]=\pm
J^\pm_{n+m},\\ {} [J^+_n,J^-_m]=-2J^3_{n+m}+kn\delta_{n+m,0}. \end{array}
\right. 
\eea
The generators $J^a_n$ can be encoded in holomorphic currents
$J^a(z)$:
\bea
J^a(z)=\sum_{n\in\Z} z^{-n-1}J^a_n \sp J^a_n = \frac{1}{2\pi i}\oint_0
dz\ z^nJ^a(z),
\eea
where $\oint_0$ stands for the integral along a contour encircling the
point $z=0$. The conformal symmetry generators $L_n$ are built from
the $\asl$ generators via the standard Virasoro construction, which
yields the central charge $c=\frac{3k}{k-2}$. 

The spectral flowed field $\Phi^{j,w}(z)$ of spin $j$ and spectral flow
number $w\in \Z$ is defined as a primary with respect to the spectral
flowed currents $\tilde{J}(z)$. These currents can be 
 defined via their modes
$\tilde{J}^a_n$, which are then used to build a spectral flowed copy
$\tilde{L}_n$ of the Virasoro algebra \cite{mo00a}:
\bea
\left\{ \begin{array}{l}
\tJ^3_n=J^3_n-\frac{k}{2}w\delta_{n,0},\\ \tJ^\pm_n=J^\pm_{n\pm w}, \\
\tilde{L}_n=L_n+wJ^3_n-\frac{k}{4} w^2\delta_{n,0}. \end{array}
\right. 
\label{flop}
\eea
Namely, the state $|j,w\rangle$ 
corresponding to the field $\Phi^{j,w}(z)$ is assumed to obey
\bea
\left\{ \begin{array}{l} \tJ^a_{n>0} |j,w\rangle =0, \\ \tJ^a_0
    |j,w\rangle = -t^a|j,w\rangle.
\end{array}\right.
\eea
Equivalently, the field $\Phi^{j,w}(z)$ has the following operator
product expansion with the {\it ordinary} currents $J^a(z)$:
\bea
\left\{ \begin{array}{l}
  J^3(z)\Phi^{j,w}(y)\sim \frac{-t^3\Phi^{j,w}(y)}{z-y}+\frac{kw}{2} \frac{\Phi^{j,w}(y)}{z-y},
\\ J^+(z)\Phi^{j,w}(y)\sim \frac{-t^+\Phi^{j,w}(y)}{(z-y)^{1+w}}, \\
J^-(z)\Phi^{j,w}(y)\sim \frac{-t^-\Phi^{j,w}(y)}{(z-y)^{1-w}}.
\end{array} \right.
\label{jope}
\eea
Here $t^a$ are generators of the $s\ell_2$ algebra. The field
$\Phi^{j,w}(z)$ indeed carries a representation of $s\ell_2$ of spin
$j$ and Casimir $\frac12(t^+t^-+t^-t^+-2t^3t^3)=-j(j+1)$,
although the corresponding degrees of freedom are not spelt out
explictly so far. I will later assume that $\Phi^{j,w}(z)$ belongs to a
principal continuous series representation with spin $j\in
-\frac12+i\R$, whose states can be labelled using a complex parameter
$\mu$ such that
\bea
\left\{ \begin{array}{l} t^+=\mu, \\ t^3 =\mu\pp{\mu} , \\ t^-
    =\mu\ppd{\mu} -\frac{j(j+1)}{\mu}. \end{array} \right. 
\label{mub}
\eea  
Another basis for the continuous representation is obtained by
diagonalizing 
$t^3$ with eigenvalue
$-m$ and considering $t^\pm$ as raising and lowering operators.
Then the state $|j,w,m\rangle$ corresponding to the field
$\Phi^{j,w}_m(z)$ satisfies
\bea
\tJ^3_0 = -t^3 = J^3_0-\frac{kw}{2}=m.
\eea
An advantage of the $m$-basis fields is that they happen to be
eigenvalues of the original dilatation operator $L_0$ and therefore
scale as follows: 
\bea
\begin{array}{c}
\left(z\pp{z} + \Delta^{j,w}_m \right) \Phi^{j,w}_m(z)=0,
\\
\Delta^{j,w}_m=\Delta_j-wm-\frac{k}{4}w^2,
\\
\Delta_j=-\frac{j(j+1)}{k-2}.
\end{array}
\label{phisc}
\eea
Moreover, the $m$-basis fields are simply related to parafermionic
fields $\Psi^j_m$ of the coset model $\SLU$ \cite{mo00a},
\bea
\Phi^{j,w}_m=e^{i\left(m+\frac{k}{2}w\right)\sqrt{\frac{2}{k}}
  \phi(z)} \Psi^j_m,
\label{paraf}
\eea
where $\phi(z)$ is a free boson such that
$J^3(z)=-i\sqrt{\frac{k}{2}}\p_z \phi(z)$. 

From the relation with parafermions (\ref{paraf}), it may seem easy to
compute the correlation function 
of $n$ spectral flowed fields and to determine the differential equations
it satisfies, by relating it to a correlation function with no
spectral flow. The only dependence on the spectral flow $w$ is indeed
in the free boson factor. However, the corresponding free boson
correlation function does make sense only if the total charge
vanishes, $\sum_{i=1}^n \left(m_i+\frac{k}{2}w_i\right)=0$. In the
case with no spectral flow $w_i=0$, this implies $\sum_{i=1}^n m_i=0$. In the
case with spectral flow, the last two equalities imply spectral flow
conservation $\sum_{i=1}^n w_i=0$. Therefore, only spectral
flow-preserving correlators are related to correlators without
spectral flow. I will use this in order to check the
KZ-type equations which I will derive for them.

\subsection{KZ-type equations for spectral flow-preserving
  correlators}

Each of the $n$ KZ equations determines the dependence of a
correlation function with respect to the worldsheet position of one
field $z_i$. This is done by inserting the worldsheet translation
operator $L_{-1}$,
\bea
L_{-1}\Phi^{j_i,w_i}(z_i) = \pp{z_i} \Phi^{j_i,w_i}(z_i).
\eea
The next step is to express $L_{-1}$ in terms of the currents $J^a$. 
Having better control on
the action of the spectral flowed currents
$\tJ^a$ on the spectral flowed field $\Phi^{j,w}(z)$, it is actually
more convenient to use
\bea
\left((k-2)\tilde{L}_{-1}+\tJ_{-1}^a\tJ^a_0 \right)
\Phi^{j_i,w_i}(z_i)=0, 
\eea
and to rely on equation (\ref{flop}) to relate $\tilde{L}_{-1}$ to the
translation operator $L_{-1}$. The resulting equation is:
\bea
\left[(k-2)\pp{z_i}
+2\tJ^3_{-1}\left(-t_i^3+\frac{k-2}{2}w_i\right)+\tJ^+_{-1}t^-_i+\tJ^-_{-1}
t^+_i \right] \Phi^{j_i,w_i}(z_i)=0.
\label{nullvect}
\eea
Now let me insert this null-vector equation into an $n$-point correlation
function. The operators $\tJ^a_{-1}$ have to be expressed in terms
of Lie algebra generators $t^a$ acting on the other fields
$\Phi^{j_\ell,w_\ell}(z_\ell)$ for $\ell\neq i$. In the case of $J^+$,
this is possible
thanks to the equation:
\bea
\la \frac{1}{2\pi i} \oint_{\infty} \frac{dz}{z-z_i}\prod_{\ell=1}^n
(z-z_\ell)^{w_\ell}\ J^+ (z)\ \prod_{\ell=1}^n
\Phi^{j_\ell,w_\ell}(z_\ell)\ra =0.
\label{zinf}
\eea
This equation holds provided
$\sum_{\ell=1}^n w_\ell \leq 0$. This indeed implies
that the function $ \prod_{\ell=1}^n
(z-z_\ell)^{w_\ell} $ is bounded near $z=\infty$ and allows
closing the contour there, knowing $J^+(z)\sim \frac{1}{z}$. 

Starting with equation (\ref{zinf}), the contour of integration can
be contracted into small loops around each point $z_\ell$. With
the help of the operator product expansion $J^a(z)\Phi^{j,w}(y)$
(\ref{jope}), this yields:
\bea
\left[\rho_i J^+_{-1,i} -\rho_i \sum_{\ell\neq i} \frac{w_\ell}{z_i-z_\ell}
t^+_i -\sum_{\ell\neq i}\frac{\rho_\ell}{z_\ell-z_i} t^+_\ell \right]
\la \prod_{\ell=1}^n \Phi^{j_\ell,w_\ell}(z_\ell) \ra =0,
\label{jpe}
\eea
where the index
$i$ in $J^+_{-1,i}$ and $t^+_i$ indicates which field they act on, and
\bea
\rho_i\equiv\prod_{\ell\neq i} (z_i-z_\ell)^{w_\ell}.
\label{rhoi}
\eea

Similar manipulations are possible with $J^-$ provided $\sum_\ell
w_{\ell} \geq 0$, and yield:
\bea
\left[\rho^{-1}_i J^-_{-1,i} +\rho^{-1}_i \sum_{\ell\neq i} \frac{w_\ell}{z_i-z_\ell}
t^-_i -\sum_{\ell\neq i}\frac{\rho^{-1}_\ell}{z_\ell-z_i} t^-_\ell \right]
\la \prod_{\ell=1}^n \Phi^{j_\ell,w_\ell}(z_\ell) \ra =0.
\label{jme}
\eea
In the case of $J^3$, the following equation does not require any
constraint on $w_\ell$:
\bea
  \left[J^3_{-1,i} -\sum_{\ell\neq
    i}\frac{1}{z_\ell-z_i}\left(t^3_\ell-\frac{k}{2}w_\ell\right)
\right] \la \prod_{\ell=1}^n \Phi^{j_\ell,w_\ell}(z_\ell) \ra =0.
\label{jte}
\eea
In the spectral flow-preserving case $\sum_{\ell=1}^n w_\ell=0$, the
three equations (\ref{jpe}),(\ref{jme}),(\ref{jte}) hold. Plugging them into equation
(\ref{nullvect}) yields the following generalization of the KZ
equations:\footnote{These equations are a special case
  of the twisted KZ equations of \cite{bho01}. Meanwhile, I believe the
  generalized KZ equations in the next subsection are new.}
\begin{multline}
\tilde{\cal E}_i \la \prod_{\ell=1}^n \Phi^{j_\ell,w_\ell}(z_\ell) \ra
=0 \ \ \ {\rm with}\ \ \ 
\tilde{\cal E}_i\equiv \left((k-2)\pp{z_i} + \sum_{j\neq
    i}\frac{\tilde{Q}_{ij}}{z_j-z_i} \right), 
\label{kzw}
\\
\tilde{Q}_{ij}\equiv  -2t^3_it^3_j+ t_i^-t_j^+ \frac{\rho_j}{\rho_i}
+  t_i^+t_j^-\frac{\rho_i}{\rho_j}
  +(k-2)(w_jt^3_i+w_it^3_j)-\frac{k(k-2)}{2}w_iw_j.
\end{multline}
These equations are related to the ordinary KZ equations ${\cal
  E}_i=0$ by a twist of
the correlation function:
\bea
{\cal E}_i \ &\kappa^{-1}& \la \prod_{\ell=1}^n
\Phi^{j_\ell,w_\ell}(z_\ell) \ra  =0,
\\
{\rm with}\ \ \ 
{\cal E}_i &\equiv& \left((k-2)\pp{z_i} + \sum_{j\neq
    i}\frac{Q_{ij}}{z_j-z_i} \right),\ \ 
Q_{ij}\equiv  -2t^3_it^3_j+ t_i^-t_j^+ 
+  t_i^+t_j^-,
\label{kz}
\\
\kappa &\equiv&  \prod_{j< i } (z_j-z_i)^{
  w_jt^3_i+w_it^3_j-\frac{k}{2} w_iw_j}.
\label{kap}
\eea
This is shown by a direct computation which uses the spacetime
$SL(2)$ invariance of the vacuum,
\bea
\la \frac{1}{2\pi i}\oint_\infty dz\ J^3(z)\ \prod_{\ell=1}^n
\Phi^{j_\ell,w_\ell}(z_\ell) \ra  =0  \ \ \Rightarrow \ \ \sum_{i=1}^n
\left(t^3_i-\frac{k}{2} w_i \right)=0 \ \ \Rightarrow \ \
\sum_{i=1}^n t^3_i =0.
\eea
A check of the equation (\ref{kz}) can be performed using the relation
of spectral flowed fields to parafermionic fields (\ref{paraf}). This
relation implies that the correlation function with spectral-flowed
fields $\la \prod_{\ell=1}^n
\Phi^{j_\ell,w_\ell}(z_\ell) \ra$ is equal to an unflowed correlator up
to free boson correlators:
\bea
\la \prod_{\ell=1}^n
\Phi^{j_\ell,w_\ell}(z_\ell) \ra = \frac{\la \prod_{\ell=1}^n
  e^{i\left(m_\ell+\frac{k}{2}w_\ell \right)\sqrt{\frac{2}{k}} 
  \phi(z_\ell)} \ra }{ \la \prod_{\ell=1}^n
  e^{im_\ell\sqrt{\frac{2}{k}} 
  \phi(z_\ell)} \ra } \la \prod_{\ell=1}^n
\Phi^{j_\ell}(z_\ell) \ra = \kappa \la \prod_{\ell=1}^n
\Phi^{j_\ell}(z_\ell) \ra,
\label{pfb}
\eea
where $m_\ell$ is by definition the eigenvalue of
$-t^3_\ell$. 

Therefore, the spectral flow-preserving correlators satisfy the
ordinary KZ equations modulo a simple twist. I will now generalize the
methods used to derive these equations to the spectral flow-violating
case. The same twist will provide notable simplifications of the
equations, without reducing them to the ordinary KZ equations.

\subsection{KZ-type equations for spectral flow-violating correlators}

Consider an $n$-point correlator which violates spectral flow by $r\geq 1$ units,
say $\sum_{\ell=1}^n w_\ell= -r$. The reasoning of the previous
subsection which led to KZ-type equations now fails because equation
(\ref{jme}), which expressed the action of $J^-_{-1}$ on a field in
terms of the action of $t^-$ on the other fields, no longer holds. To
derive such an equation would require 
\bea
 \frac{1}{2\pi i} \oint_{\infty} \frac{dz}{z-z_i}\prod_{\ell=1}^n
(z-z_\ell)^{-w_\ell}\ J^- (z)\ \overset{?}{=}0,
\label{pint}
\eea
where the l.h.s. behaves near $z=\infty $ as $ \frac{1}{2\pi i} \oint_{\infty}
dz\ z^{r-2}$. 

Actually, in the case $r=1$, the spacetime $SL(2)$ symmetry of the
vacuum is able to save the day. This symmetry indeed reads
\bea
\frac{1}{2\pi i} \oint_{\infty} dz\ J^a(z)=0,
\eea
which implies eq. (\ref{pint}). But, for $r\geq 2$, it is impossible
to derive $n$ equations governing the $z_i$ dependence of the
correlators. Instead of equation (\ref{pint}), it is however possible
to use the weaker equations:
\bea
 \frac{1}{2\pi i} \oint_{\infty} \frac{dz}{\prod_{\a=1}^{r}(z-z_{i_\a})}\prod_{\ell=1}^n
(z-z_\ell)^{-w_\ell}\ J^- (z)\ =0,
\label{weak}
\eea
for any choice of $r$ distinct indices $i_1,i_2\cdots i_r$. This leads
to an expression for a linear combination of
$J^-_{-1,i_1},J^-_{-1,i_2}\cdots J^-_{-1,i_r}$ in terms of
$t^-_\ell,\ell=1\cdots n$. Then it is possible to derive a
differential equation for the spectral flow-violating $n$-point
correlator, whose $z$-derivative part is a linear combination of
$\pp{z_{i_1}} ,\pp{z_{i_2}} ,\pp{z_{i_r}} $. It is not necessary to go
  into much detail here: these manipulations actually also hold in the
  spectral flow-preserving case, and they can therefore yield nothing
  but a linear combination of the KZ-type equations (\ref{kzw}) which
  hold in that case. The actual combination can easily be read from
  eq. (\ref{weak}),
\bea
\tilde{\cal E}_{\{i_\a\}}\equiv \sum_{\a=1}^{r} (\rho_{i_\a} t_{i_\a}^+)^{-1}
\frac{1}{\prod_{\beta\neq \alpha}(z_{i_\a}-z_{i_\beta})} \tilde{\cal E
}_{i_\a} \ \ {\rm for\ all}\ \ \{i_1,i_2\cdots
i_r\}\subset\{1,2\cdots n\},
\label{kzc}
\eea
where $\rho_i$ was defined in eq. (\ref{rhoi}). 
That only such combinations of $r$ equations hold, means that 
$r-1$ KZ equations have been lost because of spectral flow violation
$\sum_{\ell=1}^n w_\ell= -r$. This was because less
equations could be obtained from the $J^-(z)$ current. Conversely, it
is now possible to obtain new equations from the $J^+(z)$ current,
using
\bea
\oint_{\infty} dz\ \prod_{\ell=1}^n (z-z_\ell)^{w_\ell}\ z^{j} J^+(z) =0 \sp
j=0,1\cdots r.
\eea
This results in $r+1$ equations,
\bea
 \sum_{\ell=1}^n z_\ell^j \rho_\ell t^+_\ell \la \prod_{\ell=1}^n
\Phi^{j_\ell,w_\ell}(z_\ell) \ra = 0. 
\label{rpu}
\eea
The $j=0$ equation  already held in the spectral flow-preserving case
as a consequence of the spacetime $SL(2)$ symmetry. The other
equations, however, are specific to the spectral flow-violating case. 

To conclude this subsection, let me gather the equations satisfied by
the spectral flow-violating correlators, while simplifying them by
applying the twist by the function $\kappa$ (\ref{kap}),
\bea
{\cal E}_{\{i_\a\}} \  \kappa^{-1} \la \prod_{\ell=1}^n
\Phi^{j_\ell,w_\ell}(z_\ell) \ra&=&0 \ \ \ \ {\rm for} \ \ \ \{i_1,i_2\cdots
i_r\}\subset\{1,2\cdots n\},
\label{kzv}
\\
\sum_{i=1}^n z_i^j t^+_i\  \kappa^{-1} \la \prod_{\ell=1}^n
\Phi^{j_\ell,w_\ell}(z_\ell) \ra&=&0 \ \ \ \ {\rm for}\ \ \ 0\leq j\leq r,  
\label{uje}
\\
\left(\sum_{i=1}^n t^3_i +\frac{k}{2}r \right) \la \prod_{\ell=1}^n
\Phi^{j_\ell,w_\ell}(z_\ell) \ra&=&0,
\label{stt}
\eea
where ${\cal E}_{\{i_\a\}}$ is a combination of $r$ 
ordinary KZ equations ${\cal E}_i$ (\ref{kz}),
\bea
{\cal E}_{\{i_\a\}}\equiv \sum_{\a=1}^{r} (t_{i_\a}^+)^{-1}
\frac{1}{\prod_{\beta\neq \alpha}(z_{i_\a}-z_{i_\beta})} {\cal E
}_{i_\a},
\label{eia}
\eea
 and the last equation
(\ref{stt}) is the $J^3$
part of the spacetime $SL(2)$ symmetry. (The other parts are
implicitly included
in the previous equations.)

\subsection{Sklyanin's separation of variables}

The ordinary KZ equations, as well as the modified (combinations of)
KZ equations for correlators of spectral-flowed fields, involve Lie
algebra generators $t^a_i$ acting on all the fields $i=1,2\cdots n$. 
However, Sklyanin has
shown how to separate them by a change of variables
\cite{skl89}. Since it maps the KZ equations to the
Belavin--Polyakov--Zamolodchikov equations \cite{sto00}, this change
of variables leads to a relation between correlators in the Euclidean
$AdS_3$ and correlators in Liouville theory \cite{rt05}. In
preparation for the extension of such a relation to correlators of
spectral-flowed fields, I will now show how to perform the separation
of variables in the equations (\ref{kzv})-(\ref{uje}). 

I now assume that the field $\Phi^{j,w}(z)$ belongs to the principal
continuous series $j\in -\frac12+i\R$, and choose the $\mu$-basis for
this representation (see eq. (\ref{mub})). This amounts to
diagonalizing the operator $t^+$, with eigenvalue $\mu$. This could be
made explicit by using the notation
$\Phi^{j,w}(z)=\Phi^{j,w}(\mu|z)$. Then the equation
(\ref{uje}) simply becomes
\bea
u_j\equiv \sum_{\ell=1}^n \mu_\ell z_\ell^j =0 \ \ \ {\rm for}\ \ \  0\leq j\leq r.
\eea

Let me define new variables as the zeroes of the rational function
\bea
R(t)\equiv \sum_{\ell=1}^n \frac{\mu_\ell}{t-z_\ell}.
\eea
The number of zeroes of $R(t)$ is found by reducing it to the same
denominator,
\bea
R(t)=\frac{\sum_{d=0}^{n-1} \left(\sum_{j=0}^d p_j u_{d-j}\right)
  t^{n-1-d}}{\prod_{\ell=1}^n (t-z_\ell)}  \ \ {\rm where}\ \
\prod_{\ell=1}^n (t-z_\ell) = \sum_{j=0}^n p_j t^{n-j}\ \ {\rm
  defines}\ p_j.
\eea
Since $u_j=0$ for $0\leq j \leq r$, the denominator of $R(t)$ has
degree $n-2-r$. This defines the new variables $y_a$ as
\bea
\sum_{\ell=1}^n \frac{\mu_\ell}{t-z_\ell}=u_{r+1}
\frac{\prod_{a=1}^{n-2-r} (t-y_a)}{\prod_{\ell=1}^n (t-z_\ell)}.
\label{yad}
\eea
Now I am in a position to perform the change of variables
\bea
\begin{array}{c} (\mu_1, \mu_2,\cdots \mu_n)\bigl|_{u_0=u_1=\cdots
  =u_r=0} \\ (n\ {\rm variables\ subject\ to}\ r+1\ {\rm constraints})
\end{array}\ \ \ \ \rightarrow \ \ \ \ \begin{array}{c} (y_1,y_2,\cdots
  y_{n-2-r},u_{r+1}) \\ (n-r-1\ {\rm variables}) \end{array}
\label{chvar}
\eea
It is also convenient to perform a change of unknown function by
explicitly solving the equations $u_0=u_1=\cdots=u_r=0$ (\ref{uje}),
\bea
 \la \prod_{\ell=1}^n
\Phi^{j_\ell,w_\ell}(\mu_\ell|z_\ell) \ra = \kappa\
\prod_{j=0}^r \delta(u_j)\ \Omega_{n,r}\left(u_{r+1},y_1,y_2,\cdots
y_{n-2-r}|z_1,z_2,\cdots z_n\right).
\label{chf} 
\eea
\begin{claim} The
system of linear
combinations of KZ equations (\ref{kzv}) satisfied by $\kappa^{-1}\la \prod_{\ell=1}^n
\Phi^{j_\ell,w_\ell}(\mu_\ell|z_\ell) \ra$, which amounts 
to $n+1-r$ differential equations, is equivalent to $\Omega_{n,r}$
satisfying the $n-2-r$ BPZ equations characteristic of the Liouville
correlator $\la \prod_{\ell=1}^n V_{\a_\ell}(z_\ell)
\prod_{a=1}^{n-2-r} V_{-\frac{1}{2b}}(y_{a}) \ra$
\bea
\hspace{-7mm} \left[b^2\ppd{y_a} +
\sum_{a'\neq a}\biggl(\frac{1}{y_{aa'}} \pp{y_{a'}}
+\frac{\Delta_{-\frac{1}{2b}}}{y_{aa'}^2} \biggr)
+\sum_{i=1}^n\left(
\frac{1}{y_a-z_i} \pp{z_i}
+\frac{\Delta_{\a_i}}{(y_a-z_i)^2}\right)\right]
\Theta_{n,r}^{\frac{2-k}{2}} \Omega_{n,r}=0,
\label{bpz}
\eea
plus the three worldsheet $SL(2)$ equations
\bea
\sum_{i=1}^n z_i^{0,1,2} {\cal E}_i\ \Omega_{n,r} =0.
\label{wss}
\eea 
Notations: $b=(k-2)^{-\frac12},\ \ \ \a_i=b(j_i+1)+\frac{1}{2b},\ \ \ 
\Delta_\a=\a(b+b^{-1}-\a),\ \ \ 
\Delta_{-\frac{1}{2b}}=-\frac12-\frac{3}{4b^2}$,
\bea
\Theta_{n,r}\equiv \frac{\prod_{i<i'\leq n} z_{ii'} \prod_{a<a'\leq
    n-2-r} y_{aa'} }{\prod_{i=1}^n\prod_{a=1}^{n-2-r} (z_i-y_a)}.
\label{tnr}
\eea
\end{claim} 

The rest of the subsection is devoted to proving this claim.

First, notice that $\Omega_{n,r}$ satisfies  ${\cal
  E}_{\{i_\a\}}\Omega_{n,r}=0$. This follows from the equations
(\ref{kzv}), (\ref{stt})
and from
\bea
[{\cal
  E}_{\{i_\a\}},\prod_{j=0}^r \delta(u_j)]=0  \ \ \ {\rm modulo}\ \ \
\sum_{\ell=1}^n t^3_\ell +\frac{k}{2}r. 
\eea
This can be proved by a direct if tedious computation. 

Then, rewrite the equations  ${\cal
  E}_{\{i_\a\}}$ (\ref{eia}) as 
\bea
\left(\sum_{i=1}^n \frac{\prod_{\ell \neq i_1,i_2,\cdots i_r}(z_i-z_\ell)}{\prod_{a=1}^{n-2-r}
  (z_i-y_a)} {\cal E}_i\right) \Omega_{n,r}=0 \ \ \ \ {\rm for} \ \ \ \{i_1,i_2\cdots
i_r\}\subset\{1,2\cdots n\},
\label{rewr}
\eea
using
$t^+_i=\mu_i=u_{r+1}\frac{\prod_{a=1}^{n-2-r}(z_i-y_a)}{\prod_{\ell\neq
    i} (z_i-z_\ell)}$. Taking linear combinations of these equations
for different choices of $\{i_1,i_2\cdots
i_r\}$  yields the equivalent system
\bea
\left(\sum_{i=1}^n \frac{z_i^j}{\prod_{a=1}^{n-2-r}
  (z_i-y_a)} {\cal E}_i\right) \Omega_{n,r}=0,  \ \ \ \ {\rm for} \ \
\ j=0,1,\cdots n-r.
\eea
Further linear combinations of these equations lead to the worldsheet
$SL(2)$ equations (\ref{wss}) and to the equations
\bea
\sum_{i=1}^n \frac{1}{z_i-y_a} {\cal E}_i =0 \ \ \ \ {\rm for}\ \ \
a=1,2,\cdots n-2-r.
\eea 
Now these equations are equivalent to the BPZ equations
(\ref{bpz}), by the same  argument as in the spectral
flow-preserving case, see \cite{rt05}. 

As a check, I computed the $z$-scaling behaviour of $\Omega_{n,r}$
by using the equation \mbox{$\sum_{i=1}^n z_i {\cal E}_i
\Omega_{n,r}=0$}. The result agrees with the scaling
\bea
\sum_{i=1}^n z_i \pp{z_i} \ \kappa^{-1} \la \prod_{\ell=1}^n
\Phi^{j_\ell,w_\ell}(z_\ell) \ra = \left(-\sum_{i=1}^n
  \Delta_{j_i}-\frac{k}{4}r^2 \right)  \ \kappa^{-1} \la \prod_{\ell=1}^n
\Phi^{j_\ell,w_\ell}(z_\ell) \ra,
\eea
which is expected from the conformal dimensions of the operators
$\Phi^{j,w}_m$ (\ref{phisc}).

\section{Correlation functions with spectral-flowed states}

Until now I have considered general properties of conformal field
theories with a chiral $\asl$ symmetry. In this section I plan to
exploit these properties in the case of particular models. The most
physically interesting model with $\asl$ symmetry is string theory in
$AdS_3$. However, this theory has a complicated spectrum including
discrete states, and directly addressing it is difficult.
Therefore, I will consider the Euclidean version of that
model, also known as the $\H$ model. Although non unitary \cite{gaw91},
this model has the advantages of being Euclidean and of having a purely
continuous spectrum. 

My purpose is therefore to explicitly relate all correlation functions
and conformal blocks of the $\H$ model on a sphere to similar
quantities in Liouville theory, a simpler non-rational conformal field
theory. $\H$ physical correlators were already related to Liouville
theory in \cite{rt05}; now I want to extend this relation to conformal
blocks, and to correlators involving spectral-flowed fields. Such fields are
unphysical in the $\H$ model since they do not
appear in the spectrum, but they play an important r\^ole in
$AdS_3$ string theory.

In order to fully characterize $\H$ correlators, the chiral results of
the last section (namely the differential equations they satisfy) have to be
supplemented with two types of information: how the left-moving and
right-moving sectors are put together, and which structure constants
appear in the operator product expansions. These data are already known, but
in the next subsection I will recast them in a form which emphasizes
the reflection symmetry and the relation to Liouville
theory. Moreover, I will interpret them in terms of two alternative
operator product expansions in the $\H$ model.
More details on these models can be found in
\cite{rt05} and references therein.

\subsection{The $\H$ model, Liouville theory, and their structure constants}

\subsubsection{The three-point function of the $\H$ model}

The $\H$ model is a conformal field theory with symmetry algebra
$\asl\times \asl$. The spectrum is made of physical fields
$\Phi^j(z,\bz),\ j\in -\frac12+i\R$
transforming as vectors in the principal continuous series
representation of spin $j$ of both
$\asl$ algebras. Therefore, the physical fields transform as  products
$\Phi^j(z,\bz) \sim \Phi^j(z)\Phi^j(\bz)$
of the chiral fields of the previous section; however this chiral
factorization fails at the level of the zero modes \cite{gaw91}. The spectral flowed
fields $\Phi^{j,w\neq 0}(z,\bz)$ do not belong to the spectrum. 

Two different bases for the spin $j$ representation will
appear: the $\mu$-basis (see eq. (\ref{mub})) and the $m$-basis, whose elements diagonalize
the operators $t^+,\bar{t}^+ $ and $t^3,\bar{t}^3$ respectively. They
are related by 
\bea
\Phi^{j,w}_{m,\bar{m}}(z,\bz)=N^j_{m,\bar{m}} \int
\frac{d^2\mu}{|\mu|^2} \mu^{m}\bar{\mu}^{\bar{m}} \Phi^{j,w}(\mu,
\bar{\mu}|z,\bar{z})\ \  \sp\ \  N^j_{m,\bar{m}}
=\frac{\G(-j-m)}{\G(j+1+\bar{m})}.
\label{mbs}
\eea
(Note the change of convention $m\rar -m$ wrt \cite{rt05}.) In later
formulas, the antiholomorphic dependence on $\bz$ and
$\bar{\mu}$ may be omitted. The physical values of $m,\bar{m}$ obey
$m-\bar{m}\in \Z$ and $m+\bar{m}\in i\R$. 

The $\H$ two-point function has to preserve spectral flow. Therefore,
the flowed two-point function can be deduced from the unflowed one by
using formula (\ref{pfb}):
\bea
\la \Phi^{j_1,w}(\mu_1|z_1)\Phi^{j_2,-w}(\mu_2|z_2)\ra &=&
|z_{12}|^{-4\Delta_{j_1}+kw^2} |\mu_1|^2
\delta^{(2)}(\mu_1+(-1)^w z_{12}^{2w}\mu_2) 
\nonumber
\\
&\times & 
\left[
  \delta(j_2+j_1+1)+R^H(j_1)\delta(j_2-j_1)\right],  
\label{twomu}
\\
\nonumber
\la \Phi^{j_1,w}_{m_1,\bar{m}_1}(z_1)
\Phi^{j_2,-w}_{m_2,\bar{m}_2}(z_2) \ra &=&
|z_{12}|^{-4\Delta^{j_1,w}_{m_1}} (-1)^{m_1-\bar{m}_1}
\delta^{(2)}(m_1+m_2) 
\\
\times 
\bigg[
  \delta(j_2+j_1+1)&+&R^H(j_1) \frac{\G(-j_1+m_1)}{\G(j_1+1+m_1)}
\frac{\G(-j_1-\bar{m_1})}{\G(j_1+1-\bar{m_1})}\delta(j_2-j_1)\bigg],
\label{twom}
\eea
where $\Delta^{j_1,w}_{m_1}$ is defined in eq. (\ref{phisc}), and,
using the notation $b^2=\frac{1}{k-2}$, 
\bea
R^H(j)=  -
\left(\frac{1}{\pi}b^2\gamma(b^2)\right)^{-(2j+1)}
\frac{\Gamma(+2j+1)}{\Gamma(-2j-1)}
\frac{\Gamma(+b^2(2j+1))}{\Gamma(-b^2(2j+1))}. 
\eea
This $\H$ reflection coefficient is actually identical to the
Liouville theory reflection coefficient $R^L(\a)$, provided
$b^2=\frac{1}{k-2}$ is interpreted as the usual parameter of Liouville
theory (such that $c=1+6Q^2$ with $Q=b+b^{-1}$), and the Liouville
momentum $\a$ is given by \cite{sto00,rt05}
\bea
\alpha=b(j+1)+\frac{1}{2b}.
\label{ajr}
\eea

The $\H$ three-point function can either preserve spectral flow or
violate it by one unit.
Let me start by recalling the spectral flow-preserving three-point
function, while introducing new notations for the structure constants: 
\begin{multline}
\la\prod_{\ell=1}^3 \Phi^{j_\ell,w_\ell}(\mu_\ell|z_\ell)\ra
\underset{\sum w=0}{=} |z_{12}|^{-2\Delta^3_{12}-kw_1w_2}
|z_{13}|^{-2\Delta_{13}^2-kw_1w_3}
|z_{23}|^{-2\Delta^{1}_{23}-kw_2w_3}
\\
\times
\delta^{(2)}(\mu_1\rho_1+\mu_2\rho_2+\mu_3\rho_3) \
D^H\bigg[\begin{array}{ccc} j_1 & j_2 & j_3 \\ \mu_1\rho_1 & \mu_2
    \rho_2 &\mu_3 \rho_3 \end{array} \bigg] \ C^H(j_1,j_2,j_3), 
\end{multline}
with $\rho_i$ is defined by (\ref{rhoi}) and 
$\Delta^{3}_{12}=\Delta_{j_1}+\Delta_{j_2}-\Delta_{j_3}$. I define the
structure constant $C^H$ as
\begin{multline}
C^{H}(j_3,j_2,j_1)= -\frac{1}{2\pi^3 b}
\left[\frac{\gamma(b^2)b^{2-2b^2}}{\pi}\right]^{-2-\Sigma j_i} 
\frac{\up'(0)}{\up(-b(j_{123}+1)) \G(-j_{123}-1)}\label{hb3pt}
\\
\times
\frac{\up(-b(2j_1+1))\up(-b(2j_2+1))\up(-b(2j_3+1))}
{\up(-bj_{12}^3) \G(-j_{12}^3)\ \up(-bj_{13}^2) \G(-j_{13}^2)\
\up(-bj_{23}^1)\G(-j_{23}^1)},
\end{multline}
where $j_{12}^3=j_1+j_2-j_3$ and $j_{123}=j_1+j_2+j_3$ and the
definition of the special
function $\up$ can be found in \cite{rt05}. Notice the
extra $\Gamma$ factors with respect to the standard definition
\cite{tes97a,tes99}. They are added so that $C^H$ is
reflection-covariant like the three-point function itself,
\bea
C^H(j_1,j_2,j_3)=R^H(j_3)C^H(j_1,j_2,-j_3-1).
\eea
Therefore, the factor $D^H$, and the
three-point conformal block, are now reflection-invariant:
\bea
D^H\bigg[\begin{array}{ccc} j_1 & j_2 & j_3 \\ \mu_1 & \mu_2
   &\mu_3  \end{array} \bigg] &=& \pi
\left|\frac{\mu_1}{\mu_2}\right|^{2j_1+2} |\mu_2|^2\times
\label{dh}
\\
\nonumber
&\times& \sum_{\eta=\pm}\
\gamma_{j_3^\eta}^{j_1,j_2}\ 
\left|\frac{\mu_3}{\mu_2}\right|^{-2j_3^\eta}\ {}_2{\cal F}
_1\big(j_1-j_2-j_3^\eta,j_1+j_2-j_3^\eta+1,-2j_3^\eta,-\frac{\mu_3}{\mu_2}\big),
\\
\gamma_{j_3}^{j_1,j_2}&\equiv &\
\G(-j_{123}-1)\G(-j_{23}^1)\G(-j_{13}^2) \G(j^3_{12}+1)\ \gamma(2j_3+1),
\eea
where $j^+=j\ , j^-=-j-1$, and ${}_2{\cal F}
_1(a,b,c,z)=F(a,b,c,z)F(a,b,c,\bar{z})$. In this formula, the
permutation symmetry of $D^H$ is not manifest, indeed $j_3$ plays a
priviledged r\^ole and the formula could be called a
``$j_3$-decomposition'' of $D^H$ into two terms $\eta=\pm$. The other
possible decompositions associated with $j_1$ and $j_2$ naturally give
the same result; this is a consequence of the monodromy properties of
the hypergeometric function $_2F_1$, which will shortly be interpreted
as Liouville braiding.

Now let me consider the $\H$ spectral flow-violating three-point
function. This was determined in \cite{gn01,mo01}:
\begin{multline}
 \la\prod_{\ell=1}^3 \Phi^{j_\ell,w_\ell}_{m_\ell,\bar{m}_\ell}(z_\ell)\ra
\underset{\sum w=-1}{=} \left|
z_{12}^{ ^{\Delta_{m_3}^{j_3,w_3}-\Delta^{j_1,w_1}_{m_1}
  -\Delta_{m_2}^{j_2,w_2}}}
z_{13}^{ ^{\Delta_{m_2}^{j_2,w_2}-\Delta^{j_1,w_1}_{m_1}-\Delta_{m_3}^{j_3,w_3}}}
z_{23}^{ ^{\Delta^{j_1,w_1}_{m_1}-\Delta_{m_2}^{j_2,w_2}-\Delta_{m_3}^{j_3,w_3}}}
\right|^2  \\
\times \delta^{(2)}\big( \sum m_\ell -\frac{k}{2}\big)\
\prod_{\ell=1}^3 N^{j_\ell}_{m_\ell,\bar{m}_\ell} \ \times 
\tilde{C}^H(j_1,j_2,j_3),
\end{multline}
\paragraph{A useful notation.} For conciseness, the
modulus squared of $m$-dependent expressions means
\bea |Y(z,m)|^2=Y(z,m)\times Y(z\rar \bz,\ m\rar \bar{m}),
\eea
that is a product of factors depending on $m$ and $\bar{m}$, although
$m$ and $\bar{m}$ are not complex conjugates.

The spectral flow-violating structure constant has been
determined up to a $k$-dependent normalization:
\bea
\tilde{C}^H\simeq\frac{\left[\frac{1}{\pi}\gamma(b^2)b^{2-2b^2}\right]^{-j_{123}}}{
\up(b(j_{123}+1+\frac{k}{2}))}  
\frac{\up(-b(2j_1+1)) \up(-b(2j_2+1))
  \up(-b(2j_3+1))}{ \up(b(j_{12}^3+\frac{k}{2}))
  \up(b(j_{13}^2+\frac{k}{2})) \up(b(j_{23}^1+\frac{k}{2})) }.
\eea
In the $\mu$ basis, the spectral flow-violating three-point function is:
\begin{multline}
 \la\prod_{\ell=1}^3 \Phi^{j_\ell,w_\ell}(\mu_\ell|z_\ell)\ra
\underset{\sum w=-1}{=}
|z_{12}|^{\frac{k}{2}-2-kw_1w_2-2\Delta_{12}^3}
|z_{13}|^{\frac{k}{2}-2-kw_1w_3-2\Delta_{13}^2}
|z_{23}|^{\frac{k}{2}-2-kw_2w_3-2\Delta_{23}^1} 
\\
\times \frac{1}{4\pi^2}\ \delta^{(2)}\big(\sum_{\ell=1}^3 \mu_\ell\rho_\ell\big)\
\delta^{(2)}\big(\sum_{\ell=1}^3 \mu_\ell\rho_\ell z_\ell\big)\
\left|\sum_{\ell=1}^3 \mu_\ell\rho_\ell z_\ell^2\right|^{4-k}\
\tilde{C}^H(j_1,j_2,j_3).
\end{multline}
The $\mu$-dependence in this formula is derived with the
help of the results in section \ref{seckz}, in particular the equation
(\ref{rpu}). But of course the $\mu$-basis result is equivalent to
the previous $m$-basis formula.

\subsubsection{Comparison with Liouville theory}

The $\H$ structure constants $C^H(j_1,j_2,j_3)$ and
$\gamma^{j_1,j_2}_{j_3}$ are related to Liouville structure constants
with momenta $\a=b(j+1)+\frac{1}{2b}$
as follows \cite{rt05}:
\bea
\begin{array}{ccccc} C^H(j_1,j_2,j_3)\ \ \ \gamma^{j_1,j_2}_{j_3^\eta}&  &
  = &-\frac{2\pi^3}{b} &
\setlength{\fboxrule}{.2pt}
\boxed{C^L(\a_1,\a_2,\a_3+\frac{\eta}{2b})} \ \ \
\boxed{C^L_{-\eta}(\a_3)} 
\\ \psset{unit=.5cm}
\pspicture[](-3,-2.5)(3.6,3.5)
\rput[l]{120}(0,0){\extline{r}{j_1}}
\rput[l]{-120}(0,0){\extline{r}{j_2}}
\rput[l]{0}(0,0){\extline{l}{j_3}}
\rput[t]{*0}(1,-0.3){$\eta$}
\endpspicture & & = & &
\psset{unit=.5cm}
\pspicture[](-3.3,-2.5)(4.5,3.9)
\pscurve[linestyle=dashed]{->}(0,3.7)(.3,2)(.1,.5)
\pscurve[linestyle=dashed]{->}(4,3.7)(3.5,2.5)(2.5,.5)
\rput[l]{120}(0,0){\extline{r}{\a_1}}
\rput[l]{-120}(0,0){\extline{r}{\a_2}}
\rput[l]{180}(2,0){\intline{t}{\scriptstyle \a_3+\frac{\eta}{2b}}{2}}
\rput[l]{90}(2,0){\mycoil{b}{-\frac{1}{2b}}}
\rput[l]{0}(2,0){\extline{l}{\a_3}}
\endpspicture 
\end{array}
\label{fpv}
\eea
Here the diagrams illustrate relations between structure
constants, which will later be promoted into relations between
conformal blocks. These relations mean that the two terms $\eta=\pm$ of
an $\H$ vertex in the $j_3$-decomposition are equal to the two
contributions to a Liouville vertex dressed with one degenerate field
$\a=-\frac{1}{2b}$, associated with the two fusion channels $\a_3
\times -\frac{1}{2b} \rar \a_3+\frac{\eta}{2b}$. Fusing the degenerate
field with $\a_1$ or $\a_2$ would yield the $j_1$- or
$j_2$-decompositions of the $\H$ vertex respectively. The different
decompositions are therefore related by Liouville braiding as claimed
above. 

For completeness, let me recall the expressions \cite{do94,zz95} for the Liouville
structure constants which appear in eq. (\ref{fpv}): (In the relation
with the $\H$ model, the
Liouville interaction strength is fixed to $\mu_L=\frac{b^2}{\pi^2}$.)
\bea
C^{L}(\a_3,\a_2,\a_1) &=&  
\frac{\left[\pi\mu_{L}\gamma(b^2)b^{2-2b^2}
\right]^{b^{-1}(Q-\a_{123})} }{
\Upsilon_b(\alpha_{123}-Q)} 
\frac{\Upsilon_b'(0)\Upsilon_b(2\alpha_1)\Upsilon_b(2\alpha_2)
\Upsilon_b(2\alpha_3)}{\Upsilon_b(\alpha_{12}^3)
\Upsilon_b(\alpha_{23}^1)
\Upsilon_b(\alpha_{31}^2)},
\\
C^L_-(\a)&=&R^L(\a)R^L(Q-\a-\frac{1}{2b}) \sp C^L_+(\a)=1\ , 
\\
R^L(\a)&=&(\pi \mu_L 
 \gamma(b^2)  )^{\frac{Q-2\a}{b}}
\frac{\Gamma(1+b(2\a-Q))}{\Gamma(1-b(2\a-Q))}
\frac{\Gamma(1+b^{-1}(2\a-Q))}{\Gamma(1-b^{-1}(2\a-Q))}.
\eea

The comparison of the spectral flow-violating structure constant
$\tilde{C}^H$ with Liouville theory yields a very simple
result: (The authors of \cite{gn05} hint at such an equality
  but do not write it explicitly.)
\bea
\begin{array}{ccccc} \tilde{C}^H(j_1,j_2,j_3) & & = &  &
c_k\  C^L(\a_1,\a_2,\a_3) 
\\
\psset{unit=.5cm}
\pspicture[](-3,-2.5)(3.6,3)
\rput[l]{120}(0,0){\extline{r}{j_1}}
\rput[l]{-120}(0,0){\extline{r}{j_2}}
\rput[l]{0}(0,0){\extline{l}{j_3}}
\psdots[dotstyle=o,dotscale=2.3](0,0)
\endpspicture & & = & &
\psset{unit=.5cm}
\pspicture[](-3.3,-2.5)(4.5,3)
\rput[l]{120}(0,0){\extline{r}{\a_1}}
\rput[l]{-120}(0,0){\extline{r}{\a_2}}
\rput[l]{0}(0,0){\extline{l}{\a_3}}
\endpspicture 
\end{array}
\label{fpp}
\eea
Here $c_k$ is a $k$-dependent constant, whose determination would
require a more precise calculation of $\tilde{C}^H$.

\subsubsection{The operator product expansion}

Let me deduce the operator product expansion in the $\H$ model from
the two-point and three-point correlation functions. Using the
two-point function (\ref{twomu}), one can write:
\begin{multline}
\Phi^{j_1,w_1}(\mu_1|z_1) \Phi^{j_2,w_2}(\mu_2|z_2)\
\underset{z_{12}\rar 0}{\sim}\ \int_{-\frac12+i\R} dj_s\ \int
\frac{d^2\mu_s}{|\mu_s|^2}\ |z_{1s}|^{4\Delta_{j_s}-kw_s^2}
\\
\times \bigg\langle
\Phi^{j_1,w_1}(\mu_1|z_1) \Phi^{j_2,w_2}(\mu_2|z_2)
 \Phi^{-j_s-1,-w_s}(-(-1)^{w_s}z_{1s}^{-2w_s}\mu_s|z_s)\bigg\rangle
\ \ \Phi^{j_s,w_s}(\mu_s|z_1).
\label{opeb}
\end{multline}
Here, $z_s$ is an auxiliary worldsheet coordinate which disappears as
$z_{12}\rar 0$. 

It is now necessary to discuss the domain of validity of the OPE
eq. (\ref{opeb}), and in particular the value of $w_s$. When there is
no spectral flow, $w_1=w_2=0$, it is known \cite{tes99} that the OPE
holds for $w_s=0$: the OPE of unflowed fields yields only unflowed
fields. The natural generalization is: the OPE preserves
spectral flow, and eq. (\ref{opeb}) holds for $w_s=w_1+w_2$. However,
the ordinary OPE $\Phi^{j_1,0} \Phi^{j_2,0}\sim \int dj_s\ \Phi^{j_s,0}$
may not hold when the fields are applied to states outside the
physical spectrum, like states created by spectral flowed fields. For
instance, using this OPE to compute a flow-violating three-point
function $\la \Phi^{j_1,0}\Phi^{j_2,0}\Phi^{j_3,-1}\ra $ would not yield
the right result. The correct result can however be obtained
by using the OPE eq. (\ref{opeb}) with $w_s=w_1+w_2+1$. 

In string theory in the Minkowskian $AdS_3$, all values of the
spectral flow number $w_s$ appear in the spectrum and 
the OPE is expected to be of the form, $ \Phi^{j_1,w_1}
\Phi^{j_2,w_2} \sim \int dj_s\ \sum_{|w_s-w_1-w_2|\leq 1} \Phi^{j_s,w_s}$. This
cannot happen in the $\H$ model, because the spectrum is much
smaller. Indeed, according to \cite{tes99}, the $\H$ four-point
function $\la \Phi^{j_1,0}\Phi^{j_2,0}\Phi^{j_3,0}\Phi^{j_4,0}\ra $ can be
decomposed by using the ordinary, flow-preserving OPE. No extra terms
of the type $\la \Phi^{j_s,1}\Phi^{j_3,0}\Phi^{j_4,0}\ra $ appear.  

There are also cases, like the correlator  $\la
\Phi^{j_1,0}\Phi^{j_2,0}\Phi^{j_3,0}\Phi^{j_4,-1}\ra $ with one unit
of spectral flow violation, where both OPEs  $\Phi^{j_1,0}
\Phi^{j_2,0}\sim \int dj_s\ \Phi^{j_s,0}$ and   
$\Phi^{j_1,0} \Phi^{j_2,0}\sim \int dj_s\ \Phi^{j_s,1}$
can be used and yield the same result for the correlator in
question. This will be demonstrated in the next subsection, and is
evidence for the following hypothesis:
\begin{hypothesis}
The $\H$ operator product expansion eq. (\ref{opeb}) can hold with
either \mbox{$w_s=w_1+w_2-1$} or $w_s=w_1+w_2$ or $w_s=w_1+w_2+1$, depending
on which correlator the expansion is inserted in. These
possibilities are not exclusive, in particular both $w_s=w_1+w_2$ and
$w_s=w_1+w_2+1$ expansions can be used in a correlator with spectral
flow violation  $0<r<n-2$.
\end{hypothesis}
To prove this hypothesis would require a study of the fields
$\Phi^{j,w}$ as differential operators and of the domains
they act on, which I shall not attempt. Further evidence will however
come with the
result eq. (\ref{main}) for the $\H$ correlators,
derived using the hypothesis, and which is compatible with both
choices of OPEs $w_s=w_1+w_2$ and $w_s=w_1+w_2+1$, when such a choice
is available. 

Moreover, the hypothesis is consistent with the definition of spectral
flowed correlators reported in \cite{mo01,gn05} and attributed to
Fateev, Zamolodchikov and Zamolodchikov. This definition indeed involves the
insertion of the $\H$ operator of spin $-\frac{k}{2}$:
\bea
\hspace{-7mm} \la \Phi^{j_1,1}(z_1)\Phi^{j_2,0}(z_2)\cdots \ra \propto \underset{u\rar
  z_1}{\lim} \la
\Phi^{j_1}(z_1)\Phi^{j_2}(z_2)\Phi^{-\frac{k}{2}}(u)\cdots \ra \propto \la
\Phi^{-j_1-\frac{k}{2}}(z_1) \Phi^{j_2}(z_2)\cdots \ra.
\eea
In the limit $z_1\rar z_2$, this leads to the flow-violating OPE:
\bea
\underset{z_1\rar z_2}{\lim} \la
\Phi^{j_1,1}(z_1)\Phi^{j_2,0}(z_2)\cdots \ra  \propto
\underset{z_1\rar z_2}{\lim}  \la
\Phi^{-j_1-\frac{k}{2}}(z_1) \Phi^{j_2}(z_2)\cdots \ra \propto 
\int dj_s\ \la \Phi^{j_s,0} \cdots \ra , 
\eea
where the last operation was an ordinary $\H$ OPE which yielded an
unflowed field. However inverting the limits $z_1\rar z_2 $ and
$u\rar z_1$ leads to
\begin{multline}
\underset{u\rar
  z_1}{\lim} \underset{z_1\rar z_2}{\lim}   \la
\Phi^{j_1}(z_1)\Phi^{j_2}(z_2)\Phi^{-\frac{k}{2}}(u)\cdots \ra \propto \underset{u\rar
  z_1}{\lim} \int dj_s\ \la \Phi^{j_s}(z_1)
\Phi^{-\frac{k}{2}}(u)\cdots \ra 
\\
\propto \int dj_s\ \la
\Phi^{j_s,1}(z_1)\cdots \ra , 
\end{multline}
i.e. a spectral flow-preserving OPE. 

The hypothesis above suggests that the states $|j,w\neq 0\rangle$
created by spectral-flowed operators in the $\H$ model are similar to
the states $|j\not\in -\half+i\R,0\rangle$: they do not belong to the
physical spectrum and do not appear in the physical OPE, but they have
a non-vanishing three-point function with physical states, which can
be accounted for by a non-physical OPE. However, in contrast to the
flow-violating OPE, the non-phyisical OPE
involving $j\not\in -\half+i\R$ is obtained from the physical OPE by
deforming the contour of integration $\int_{-\half+i\R}dj_s$.

Now here are explicit expressions for the OPEs derived from
eq. (\ref{opeb}), obtained by inserting the explicit expression for
the three-point function.
This yields 
the spectral flow-preserving OPE in the case $w_s=w_1+w_2$,
\begin{multline}
\Phi^{j_1,w_1}(\mu_1|z_1)\Phi^{j_2,w_2}(\mu_2|z_2)\ \underset{z_{12}\rar
  0}{\sim}\
  \int_{-\frac12+i\R} dj_s\ C^H(j_1,j_2,-j_s-1)
|z_{12}|^{-2\Delta_{12}^s-kw_1w_2}
\\ \times 
|\mu_s|^{-2} D^H\left[\begin{array}{ccc} j_1 & j_2 & -j_s-1 \\ \mu_1
    z_{12}^{w_2} & \mu_2 z_{21}^{w_1} & -\mu_s \end{array} \right]\ 
\Phi^{j_s,w_1+w_2}(\mu_s=\mu_1z_{12}^{w_2}+\mu_2z_{21}^{w_1}|z_1),
\label{opep}
\end{multline}
and the spectral flow-violating OPE in the case $w_s=w_1+w_2+1$,
\begin{multline}
\Phi^{j_1,w_1}(\mu_1|z_1)\Phi^{j_2,w_2}(\mu_2|z_2) \ \underset{z_{12}\rar
  0}{\sim}\
 \frac{1}{4\pi^2} \delta^{(2)}\big(\mu_1
z_{12}^{w_2+1} -\mu_2 z_{21}^{w_1+1}\big) \left|\mu_1
  z_{12}^{w_2+1}\right|^{2-k}
\\
\times  \int_{-\frac12+i\R} dj_s\ 
 \tilde{C}^H(j_1,j_2,-j_s-1) 
|z_{12}|^{-2\Delta_{12}^s+\frac{k}{2}-kw_1w_2} \ 
\Phi^{j_s,w_1+w_2+1}(\mu_1 z_{12}^{w_2+1}|z_1).
\label{opev}
\end{multline}

\subsection{$\H$ correlators from Liouville theory}

\subsubsection{Results}

The relations between the $\H$ model and Liouville theory at the levels
of structure constants eq. (\ref{fpv}),
(\ref{fpp}), and differential equations reflecting chiral symmetry
eq. (\ref{bpz}), lead to the following expression for the $\H$
correlation functions: 
\bea
\boxed{ \begin{aligned} & \la \prod_{\ell=1}^n
    \Phi^{j_\ell,w_\ell}(\mu_\ell|z_\ell)\ra\ \ 
\underset{ {\scriptstyle \sum w = -r \leq 0}}{=} \ \ \ 
\frac{\pi}{2}(-\pi)^{-n}b\ c_k^r\ \times 
\\ & \prod_{j=0}^r
\delta^{(2)}\big({\textstyle \sum} \mu_\ell \rho_\ell z_\ell^j\big)\
\left|{\textstyle \sum}
  \mu_\ell \rho_\ell z_\ell^{r+1}\right|^{2+2r-kr}\
 |\Theta_{n,r}|^{k-2} \ \la \prod_{\ell=1}^n V_{\a_\ell}(z_\ell)
\prod_{a=1}^{n-2-r} V_{-\frac{1}{2b}}(y_{a}) \ra
\end{aligned}
}
\label{main}
\eea
Let me recall the notations involved in this formula: $V_\a(z)$ is the
Liouville vertex operator of conformal weight $\Delta_{\a}=\a(b+b^{-1}-\a)$,
where the Liouville parameter is $b=(k-2)^{-\frac12}$ and the
interaction strength is $\mu_L=b^2/\pi^2$; the Liouville momenta
$\a=b(j+1)+\frac{1}{2b}$ are such that
$\Delta_\a=\Delta_j+\frac{k}{4}$; and
the positions of the
Liouville degenerate fields $y_a$ are defined by
\bea
\sum_{\ell=1}^n \frac{\mu_\ell\rho_\ell }{t-z_\ell}= \left(\tsum_{\ell=1}^n
  \mu_\ell \rho_\ell z_\ell^{r+1} \right)
\frac{\prod_{a=1}^{n-2-r} (t-y_a)}{\prod_{\ell=1}^n (t-z_\ell)} \ \ \
\ \ 
{\rm with} \ \ \ \ \ 
\rho_\ell=\prod_{j\neq \ell}z_{\ell j}^{w_{j}}.
\eea
The factor $\Theta_{n,r}$ was
defined in eq. (\ref{tnr}), and the $k$-dependent factor $c_k$ is not
known. ($c_k$ is related but not equal to the $c_k$ of  
eq. (\ref{fpp}); other unknown $c_k$s will appear below.)

The $m$-basis $\H$ correlators are related to Liouville correlators by
applying the change of basis (\ref{mbs}) and the change of variables
(\ref{chvar}), whose Jacobian is:
\bea
 \prod_{i=1}^n \frac{d^2\mu_i}{|\mu_i|^2} \delta^{(2)}({\textstyle \sum} \mu_iz_i^r) \cdots
 \delta^{(2)}({\textstyle \sum}\mu_i )  = \frac{d^2u}{|u|^{4+2r}}
 \prod_{a=1}^{n-2-r} d^2y_a\ \frac{\prod_{a<a'}|y_{aa'}|^2
   \prod_{i<i'} |z_{ii'}|^2}{\prod_i\prod_a |y_a-z_i|^2}.
\eea
The result is:
\bea
\boxed{ \begin{aligned}
& \la \prod_{\ell=1}^n
    \Phi^{j_\ell,w_\ell}_{m_\ell,\bar{m}_\ell}(z_\ell)\ra \ 
\underset{ {\scriptstyle \sum w = -r \leq 0}}{=} \ \ \ 
\frac{2\pi^{3-2n}b\ c_k^r}{(n-2-r)!}\ 
\prod_{\ell=1}^n N^{j_\ell}_{m_\ell,\bar{m}_\ell}\
 \times \delta^{(2)}\big(\tsum m_\ell -\frac{k}{2}r\big) 
\\
& \times \left|\prod_{\ell<\ell'} z_{\ell\ell'}^{ ^{\beta_{\ell\ell'}}}\right|^2 
\int \prod_{a=1}^{n-2-r} d^2y_a\
  \frac{\prod_{a<a'}|y_{aa'}|^k}{\left|\prod_{\ell,a}
      (z_\ell-y_a)^{\frac{k}{2}-m_\ell}  \right|^2} 
\la \prod_{\ell=1}^n V_{\a_\ell}(z_\ell)
\prod_{a=1}^{n-2-r} V_{-\frac{1}{2b}}(y_{a}) \ra
\end{aligned} }
\label{mmmm}
\eea
where the combinatorial factor $\frac{1}{(n-2-r)!}$ comes from the
invariance of the $\mu_\ell$ wrt permutations of the $y_a$s, and the
exponent $\beta_{\ell\ell'}$ is defined by
\bea
\beta_{i \ell }\equiv \frac{k}{2}-\frac{k}{2}w_i w_{\ell} -w_i m_\ell
-w_\ell m_i -m_i-m_\ell.
\label{bil}
\eea
The formula (\ref{mmmm}) can be rephrased in the language of the
parafermions (\ref{paraf}) and it gives the $n$-point function $\la
\prod_{\ell=1}^n \Psi^{j_\ell}_{m_\ell,\bar{m}_\ell}\ra$ provided
$\beta_{i\ell}$ is replaced with
\bea
\beta'_{i\ell}=\frac{2}{k}(m_i-\frac{k}{2})(m_\ell-\frac{k}{2}).
\eea
The resulting expression for the parafermionic correlators agrees with the
unpublished results of Fateev \cite{fat96}, obtained by free field methods. 

The integrals over $y_a$ in eq. (\ref{mmmm}) may have singularities at
$y_a=z_i$, depending on the values of $m_\ell,\bar{m}_\ell$. The
physical values in the $\H$ model are $m-\bar{m}\in \Z,\ m+\bar{m}\in
i\R$ and would make the integrals converge, but they 
are forbidden by the constraints $\sum m=\sum
\bar{m}=\frac{k}{2}r$. However, assuming the external spins are
physical $j_\ell\in -\half+i\R$,
the integrals actually converge
provided $\Re(m_\ell+\bar{m}_\ell) > -1$. This mild condition
from the point of view of the $\H$ model becomes a problem when
Wick-rotating to string theory in $AdS_3$, whose physical spectrum
satisfies $m_\ell+\bar{m}_\ell \in \R$.

\subsubsection{Arguments}

Here I argue for (and partly prove) the relation between $\H$ correlators in the
$\mu$ basis and Liouville theory correlators, eq. (\ref{main}). In the
case when there is no spectral flow $w_\ell=0$, this has been done in
\cite{rt05}. Then, in the spectral flow-preserving case $\sum w_\ell =
-r=0$, this is a simple consequence of the formula (\ref{pfb}), where
the action of the differential operator
$\kappa$ (\ref{kap}) on $\mu_\ell$
accounts for the $\rho_\ell$ factors. 

Now in the general case, it is possible to use
an argument similar to the one in \cite{rt05} in order to reduce the
problem to maximally flow-violating correlators: whenever spectral
flow violation is not maximal, it is possible to reduce the number of
fields by using the
flow-preserving OPE and the correspondence
between KZ and BPZ equations of section \ref{seckz}.

However, I can only conjecture the validity of eq. (\ref{main}) in the
maximally flow-violating case. This is because the corresponding
Liouville correlator involves no degenerate field and thus satisfies
no BPZ equation. The $z$-dependence of that Liouville correlator is
therefore not controlled by the techniques used so far. Nevertheless,
I will present some strong evidence in favour of that
conjecture. This comes in addition to the calculations of Fateev
\cite{fat96} that eq. (\ref{mmmm}) holds whenever both sides are
accessible to free field computations. 
 
First, the proposed relation with
Liouville theory eq. (\ref{main}) is compatible with the spectral
flow-violating OPE eq. (\ref{opev}). The comparison between the
spectral flow-violating structure constant $\tilde{C}^H$ and Liouville
theory (\ref{fpp}) shows that the OPE coefficients agree. Now consider
the quantities $u_j\equiv \sum_{\ell=0}^n \mu_\ell \rho_\ell z_\ell^j$
appearing in an $n$-point function, and
\bea
u'_j \equiv \mu_1 z_{12}^{w_2+1} \prod_{\ell \geq 3}
z_{1\ell}^{w_\ell} + \sum_{i\geq 3} \mu_i z_{i1}^{w_1+w_2+1}
\prod_{\underset{\ell\neq i}{\ell \geq 3}} z_{i\ell}^{w_\ell},
\eea
which appears in the $n-1$-point function obtained by a spectral
flow-violating OPE. Direct computation leads to
\bea
u'_j \ \underset{z_{12}\rar 0}{\sim}\ u_{j+1}-z_1 u_j.
\eea
This equation is the key to showing that the positions $y_a$ of the
$n-2-r$ 
auxiliary fields are not affected by the OPE, and that the
delta-function factors behave correctly:
\begin{multline}
\delta^{(2)}\big(\mu_1 z_{12}^{w_2+1}-\mu_2z_{21}^{w_1+1}\big)\ \prod_{j=0}^{r-1}
\delta^{(2)}(u'_j)  \\  \underset{z_{12}\rar 0}{\sim}\ \delta^{(2)}(z_{12}u_0)\
\prod_{j=0}^{r-1} \delta^{(2)}(u_{j+1}-z_1 u_j ) 
= |z_{12}|^{-2}
\prod_{j=0}^r \delta^{(2)}(u_j).
\end{multline}
These manipulations with $\delta$-functions will not be very rigorous as
long as I do not define the domain of these distributions. An
alternative is to prove the equivalent $m$-basis result (\ref{mmmm})
instead of the $\mu$-basis result. This is actually quite
straightforward, but I have given the argument in the $\mu$-basis in
order to illustrate the $\mu$-basis OPE. One reason to insist on the
use of the $\mu$-basis is that conformal blocks in this basis are much
simpler than in any other basis, as will be demonstrated in the next subsection.

The second line of evidence in favour of the relation (\ref{main}) in
the maximally flow-violating case comes from the study of the special
values of $\alpha$ and $j$ where the correlators satisfy differential
equations. On the Liouville side, this happens if $\alpha$ belongs to
the Kac table associated with the central charge $c=1+6Q^2$. The
corresponding values of $j$ turn out to give $\asl$
representations with affine null vectors. Further evidence could be
obtained by comparing the null vector differential equations themselves, but this
is beyond the scope of the present article.

\subsection{$\H$ conformal blocks from Liouville theory}

The relation (\ref{main}) between $\H$ correlators and Liouville
correlators can be decomposed into relations between the structure
constants of the theories, which I already wrote (\ref{fpv}),
(\ref{fpp}), and relations between the conformal blocks. 

Consider an $n$-point correlator in $\H$ with $r$ units of spectral
flow violation, $\sum w=-r\leq 0$. I will consider decomposition
in conformal blocks which use vertices with winding violation $0$ or
$-1$. If vertices with winding violation $+1$ were included, there
would probably be no simple relation to Liouville theory. Moreover,
for ease of writing I will only consider a specific case
\mbox{$n=6,r=2$}, which involves $n-2-r=2$ Liouville degenerate field insertions.

A basis of  $\H$ non-chiral
conformal blocks is defined as follows:
\begin{multline}
\la \prod_{\ell=1}^6 \Phi^{j_\ell,w_\ell}(\mu_\ell|z_\ell) \ra = \int_{-\half+i\R}
dj_{12}\ dj_{34}\ dj_{56}\ \tilde{C}^H(j_1,j_2,-j_{12}-1)\ C^H(j_3,j_4,-j_{34}-1)
\\ \times C^H(j_5,j_6,-j_{56}-1)\
\tilde{C}^H(j_{12},j_{34},j_{56})\ 
\psset{unit=.4cm}
\pspicture[](-5,-2.3)(5,2.7)
\rput[l]{110}(-3,0){\extline{r}{2}}
\rput[l]{-110}(-3,0){\extline{r}{1}}
\rput[l]{70}(3,0){\extline{l}{5}}
\rput[l]{-70}(3,0){\extline{l}{6}}
\rput[l]{20}(0,2){\extline{l}{4}}
\rput[l]{160}(0,2){\extline{r}{3}}
\rput[l]{180}(0,0){\intline{t}{j_{12}}{3}}
\rput[l]{180}(3,0){\intline{t}{j_{56}}{3}}
\rput[l]{90}(0,0){\intline{r}{j_{34}}{2}}
\psdots[dotstyle=o,dotscale=2](-3,0)
\psdots[dotstyle=o,dotscale=2](0,0)
\endpspicture\ .
\end{multline}
The pictorial representation for the conformal block leaves its
dependence on $\mu_\ell,z_\ell,j_\ell,w_\ell$ implicit. Note that with our
definition of the structure constant $C^H$ (\ref{hb3pt}), the
conformal block is invariant wrt reflection of external spins
$j_1\cdots j_6$ and internal spins $j_{12},j_{34},j_{56}$. 

This $\H$ non-chiral conformal block can now be decomposed into
Liouville chiral conformal blocks in the following way:
\begin{multline}
\label{cbl}
\psset{unit=.4cm}
\pspicture[](-5,-2)(5,2.7)
\rput[l]{110}(-3,0){\extline{r}{2}}
\rput[l]{-110}(-3,0){\extline{r}{1}}
\rput[l]{70}(3,0){\extline{l}{5}}
\rput[l]{-70}(3,0){\extline{l}{6}}
\rput[l]{20}(0,2){\extline{l}{4}}
\rput[l]{160}(0,2){\extline{r}{3}}
\rput[l]{180}(0,0){\intline{t}{j_{12}}{3}}
\rput[l]{180}(3,0){\intline{t}{j_{56}}{3}}
\rput[l]{90}(0,0){\intline{r}{j_{34}}{2}}
\psdots[dotstyle=o,dotscale=2](-3,0)
\psdots[dotstyle=o,dotscale=2](0,0)
\endpspicture
=\frac{\pi}{2}(-\pi)^{-n}b\ c_k^r\ 
 \prod_{j=0}^r
\delta^{(2)}\big({\textstyle \sum} \mu_\ell \rho_\ell z_\ell^j\big)\
\left|{\textstyle \sum}
  \mu_\ell \rho_\ell z_\ell^{r+1}\right|^{2+2r-kr}\
\\ 
\times |\Theta_{n,r}|^{k-2} \ 
\sum_{\eta_4=\pm} \gamma^{j_3,j_{34}}_{j_4^{\eta_4}}\
\sum_{\eta_{56}=\pm} \gamma^{j_5,j_6}_{j_{56}^{\eta_{56}}}\ 
\left|
\psset{unit=.4cm}
\pspicture[](-5,-2.3)(5,3)
\rput[l]{110}(-3,0){\extline{r}{\a_2}}
\rput[l]{-110}(-3,0){\extline{r}{\a_1}}
\rput[l]{70}(3,0){\extline{l}{\a_5}}
\rput[l]{-70}(3,0){\extline{l}{\a_6}}
\rput[l]{20}(0,2){\extline{l}{\a_4}}
\rput[l]{160}(0,2){\extline{r}{\a_3}}
\rput[l]{180}(0,0){\intline{t}{\a_{12}}{3}}
\rput[l]{180}(2,0){\intline{t}{\a_{56}}{2}}
\rput[l]{180}(3,0){\intline{t}{}{1}}
\rput[l]{90}(0,0){\intline{r}{\a_{34}}{2}}
\rput[l]{90}(2,-2.2){\mycoil{b}{\eta_{56}}}
\rput[l]{-30}(-1.2,3.5){\mycoil{tl}{\eta_4}}
\endpspicture
\right|^2\ ,
\end{multline}
where the indices $\eta_4,\eta_{56}$ indicate the fusion channels of
the two degenerate fields, $\a_4+\frac{\eta_4}{2b}$ and
$\a_{56}+\frac{\eta_{56}}{2b}$, and the $z_\ell,y_a$ dependence of the
Liouville conformal block on the positions of the fields is omitted. 
Alternative positionings of the
degenerate field insertions are possible, as long as they remain
around the Liouville vertices which correspond to the spectral flow-preserving
vertices in $\H$. 
Naturally, $\H$ conformal blocks in the $m$-basis (and thus
parafermionic conformal blocks) can also be expressed in terms of
Liouville theory conformal blocks in a similar manner:
\begin{multline}
\psset{unit=.4cm}
\pspicture[](-5,-2)(5,2.7)
\rput[l]{110}(-3,0){\extline{r}{2}}
\rput[l]{-110}(-3,0){\extline{r}{1}}
\rput[l]{70}(3,0){\extline{l}{5}}
\rput[l]{-70}(3,0){\extline{l}{6}}
\rput[l]{20}(0,2){\extline{l}{4}}
\rput[l]{160}(0,2){\extline{r}{3}}
\rput[l]{180}(0,0){\intline{t}{j_{12}}{3}}
\rput[l]{180}(3,0){\intline{t}{j_{56}}{3}}
\rput[l]{90}(0,0){\intline{r}{j_{34}}{2}}
\psdots[dotstyle=o,dotscale=2](-3,0)
\psdots[dotstyle=o,dotscale=2](0,0)
\rput[b]{0}(0,-2.6){($m$-{\rm basis})}
\endpspicture
=
\frac{2\pi^{3-2n}b\ c_k^r}{(n-2-r)!}\ 
\prod_{\ell=1}^n N^{j_\ell}_{m_\ell,\bar{m}_\ell}\
 \times \delta^{(2)}\big(\tsum m_\ell -\frac{k}{2}r\big) 
\
\left|\prod_{\ell<\ell'} z_{\ell\ell'}^{ ^{\beta_{\ell\ell'}}}\right|^2 
\\ \times  \sum_{\eta_4=\pm}
  \gamma^{j_3,j_{34}}_{j_4^{\eta_4}}\ 
\sum_{\eta_{56}=\pm} \gamma^{j_5,j_6}_{j_{56}^{\eta_{56}}}\ 
\int \prod_{a=1}^{n-2-r} d^2y_a\
  \frac{\prod_{a<a'}|y_{aa'}|^k}{\left|\prod_{\ell,a}
      (z_\ell-y_a)^{\frac{k}{2}-m_\ell}  \right|^2} 
\left|
\psset{unit=.4cm}
\pspicture[](-5,-2.3)(5,3)
\rput[l]{110}(-3,0){\extline{r}{\a_2}}
\rput[l]{-110}(-3,0){\extline{r}{\a_1}}
\rput[l]{70}(3,0){\extline{l}{\a_5}}
\rput[l]{-70}(3,0){\extline{l}{\a_6}}
\rput[l]{20}(0,2){\extline{l}{\a_4}}
\rput[l]{160}(0,2){\extline{r}{\a_3}}
\rput[l]{180}(0,0){\intline{t}{\a_{12}}{3}}
\rput[l]{180}(2,0){\intline{t}{\a_{56}}{2}}
\rput[l]{180}(3,0){\intline{t}{}{1}}
\rput[l]{90}(0,0){\intline{r}{\a_{34}}{2}}
\rput[l]{90}(2,-2.2){\mycoil{b}{\eta_{56}}}
\rput[l]{-30}(-1.2,3.5){\mycoil{tl}{\eta_4}}
\endpspicture
\right|^2\ .
\end{multline}
What is however unique to the $\mu$-basis is the possibility of 
explicitly writing the $\H$ conformal blocks in some limits, obtained by performing
multiple OPEs (\ref{opep}), (\ref{opev}), for instance
\begin{multline}
\psset{unit=.4cm}
\pspicture[](-5,-2.8)(5,2.7)
\rput[l]{110}(-3,0){\extline{r}{2}}
\rput[l]{-110}(-3,0){\extline{r}{1}}
\rput[l]{70}(3,0){\extline{l}{5}}
\rput[l]{-70}(3,0){\extline{l}{6}}
\rput[l]{20}(0,2){\extline{l}{4}}
\rput[l]{160}(0,2){\extline{r}{3}}
\rput[l]{180}(0,0){\intline{t}{j_{12}}{3}}
\rput[l]{180}(3,0){\intline{t}{j_{56}}{3}}
\rput[l]{90}(0,0){\intline{r}{j_{34}}{2}}
\psdots[dotstyle=o,dotscale=2](-3,0)
\psdots[dotstyle=o,dotscale=2](0,0)
\endpspicture 
\underset{z_{12},z_{34}\ll z_{13} \ll z_{15}\ll
  z_{56}}{\sim}\ \ \ \frac{1}{16\pi^4}\  
\delta^{(2)}(\mu_1 z_{12}^{w_2+1}-\mu_2 z_{21}^{w_1+1})
\\
\times \delta^{(2)}(\mu_{12}z_{13}^{w_3+w_4+1} -\mu_{34}z_{31}^{w_1+w_2+2})\
\delta^{(2)}(\mu_{12}z_{13}^{w_3+w_4+1} z_{15}^{w_5}+ \mu_5
z_{51}^{-w_5-w_6} +\mu_6 z_{61}^{-2w_6}) \\ 
\times |\mu_{12}|^{2-2k}
|z_{12}|^{-2\Delta_{1,2}^{12}+\frac{k}{2}-kw_1w_2}
|z_{34}|^{-2\Delta_{3,4}^{34}-kw_3w_4}|z_{16}|^{-4\Delta_6+kw_6^2}
\\
\times
|z_{13}|^{-2\Delta_{12,34}^{56}+4+2(w_1+w_2)-k(w_3+w_4+1) +\frac{k}{2}-k(w_3+w_4)(w_1+w_2+1) }
|z_{15}|^{-2\Delta_{56,5}^6+kw_5(w_5+w_6)} 
\\
\times
D^H\left[\begin{array}{ccc} j_3 & j_4 & j_{34} \\ \mu_3
    z_{34}^{w_4}\ &\ \mu_4 z_{43}^{w_3}\ &\ -\mu_{34} \end{array} \right]
\
D^H\left[\begin{array}{ccc} j_{56} & j_5 & j_6 \\
    \mu_{12}z_{13}^{w_3+w_4+1} z_{15}^{w_5}\ &\ \mu_5 z_{51}^{-w_5-w_6}\
    &\ \mu_6 z_{61}^{-2w_6} \end{array} \right]\ ,
\end{multline}
where $\mu_{34}=\mu_3z_{34}^{w_4}+\mu_4z_{43}^{w_3}$ and
$\mu_{12}=\mu_1 z_{12}^{w_2+1}$. In the case where all spectral flow
numbers $w_\ell$ vanish, the conformal blocks reduce in such limits to
$SL(2,\C)$ coinvariants which have an explicit expression as products
of the $D^H$ coefficients:
\begin{multline}
\psset{unit=.4cm}
\pspicture[](-5,-3.1)(5,2.7)
\rput[l]{110}(-3,0){\extline{r}{2}}
\rput[l]{-110}(-3,0){\extline{r}{1}}
\rput[l]{70}(3,0){\extline{l}{5}}
\rput[l]{-70}(3,0){\extline{l}{6}}
\rput[l]{20}(0,2){\extline{l}{4}}
\rput[l]{160}(0,2){\extline{r}{3}}
\rput[l]{180}(0,0){\intline{t}{j_{12}}{3}}
\rput[l]{180}(3,0){\intline{t}{j_{56}}{3}}
\rput[l]{90}(0,0){\intline{r}{j_{34}}{2}}
\rput[b]{0}(0,-2.6){($w_\ell=0$)} 
\endpspicture 
\underset{z_{12},z_{34}\ll z_{13} \ll z_{15}\ll
  z_{56}}{\sim}\ \\ 
|z_{12}|^{-2\Delta_{1,2}^{12}} |z_{34}|^{-2\Delta_{3,4}^{34}}
|z_{13}|^{-2\Delta_{12,34}^{56}} |z_{15}|^{-2\Delta_{56,5}^{6}} |z_{16}|^{-4\Delta_6}\
\delta^{(2)}(\tsum \mu_\ell)\ |\mu_{12}|^{-2}
|\mu_{34}|^{-2} |\mu_{56}|^{-2}
\\ \times
D^H\left[\begin{array}{ccc} j_1 & j_2
    & j_{12} \\ \mu_1 & \mu_2 & -\mu_{12} \end{array}\right] \
D^H\left[ \begin{array}{ccc} j_3 & j_4 & j_{34} \\ \mu_3 &\mu_4&
   - \mu_{34} \end{array} \right] \ 
D^H\left[\begin{array}{ccc} j_5 & j_6 & j_{56} \\ \mu_5 & \mu_6 &-
    \mu_{56} \end{array} \right]\ 
D^H\left[\begin{array}{ccc} j_{12} & j_{34} & j_{56} \\ \mu_{12} &
    \mu_{34} & \mu_{56} \end{array} \right]\ ,
\end{multline}
where $\mu_{\ell \ell'}=\mu_\ell+\mu_{\ell'}$. The $\H$
flow-preserving conformal
blocks are fully determined by their behaviour in such limits, plus the
KZ equations. 

\section{Outlook}

The present results may hopefully be useful in the study of string
theory in $AdS_3$. Correlators in this theory should be obtained from
$\H$ correlators in the $m$ basis (\ref{mmmm}) by Wick-rotation of
$m_\ell+\bar{m}_\ell$, in the spirit of \cite{mo01}. 
The integrals in eq. (\ref{mmmm}) then become
divergent, and regularizing them should lead to the appearance of
discrete states in the intermediate channels, even if all external
states are continuous.

The expression (\ref{cbl}) of $\H$ conformal blocks in terms of
Liouville theory conformal blocks suggests a way to define and compute
the fusing matrix in the $\H$ model. This fusing matrix is still not well
understood, see \cite{pon02}. However, knowing all $\H$ correlators in
terms of Liouville correlators makes this issue less crucial for the $\H$
model proper. Nevertheless, an $\H$ fusing matrix should be a
very interesting object in itself, which may have an interpretation in
terms of harmonic analysis on the quantum group $U_q(s\ell_2)$.

\acknowledgments{I am grateful to Thomas Quella, Andreas Recknagel and
  G\'erard Watts for interesting conversations. Some ideas in this
  work arose while collaborating with Joerg Teschner on related
  issues, and I also thank him for comments on this manuscript.
  Moreover, I wish to thank Vladimir Fateev for a copy of his
  unpublished note \cite{fat96}.

  I am supported by the EUCLID European network,
  contract number HPRN-CT-2002-00325, and also in part by the PPARC
  rolling grant PPA/G/O/2002/00475. 
}





\providecommand{\href}[2]{#2}\begingroup\raggedright\endgroup


\end{document}